\documentclass[11pt]{article}
\usepackage{cospar}
\usepackage{url}
\usepackage{graphicx}
\newcommand{\rthz}{$/\sqrt{\mathrm{Hz}}$}
\newcommand{\figr}[1]{Fig.~\ref{#1}}
\newcommand{\eqr}[1]{Eq.~\ref{#1}}

\newcommand{\wsq}[1]{\omega^2_{#1}}
\newcommand{\Sh}[1]{\ensuremath{S_{#1}^{1/2}}}
\newcommand{\x}{{\it x}} 
\newcommand{\y}{{\it y}} 
\newcommand{\z}{{\it z}}
\newcommand{\wm}{\ensuremath{\omega_0}}
\newcommand{\dv}[1]{\ensuremath{\delta V_{#1}}}
\newcommand{\D}[1]{\ensuremath{\Delta _{#1}}}
\newcommand{\cder}[2]{\ensuremath{\frac{\partial C_{#1}}{\partial {#2}}}}
\newcommand{\cdder}[2]{\ensuremath{\frac{\partial ^2 C_{#1}}{\partial {#2}^2}}}
\newcommand{\dcdxdphi}
{\ensuremath{\frac{\partial ^2 C_x}{\partial x \partial \phi}}}

\newcommand{\etal}{et al.}

\title{Possibilities for Measurement and Compensation of 
Stray DC Electric Fields Acting on Drag-Free Test Masses }

\author{W.~J.~Weber\address{Dipartimento di Fisica and INFN,
Universit\`{a} di Trento, 38050 Povo (TN), Italy}, L.~Carbone$^1$,
A.~Cavalleri\address{Centro Fisica degli Stati Aggregati, 38050
Povo (TN), Italy}, R.~Dolesi$^1$, C.~D.~Hoyle$^1$, M.~Hueller$^1$, and
S.~Vitale$^1$}

\begin{document}
\maketitle

\begin{abstract}
DC electric fields can combine with test mass charging and thermal
dielectric voltage noise to create significant force noise acting on the
drag-free test masses in the LISA (Laser Interferometer Space Antenna)
gravitational wave mission. This paper proposes a simple technique to
measure and compensate average stray DC potentials 
at the mV level, yielding substantial reduction in this source
of force noise. We discuss the attainable resolution for both flight
and ground based experiments.
\end{abstract}

\section{INTRODUCTION}
In the envisioned design of the gravitational wave mission LISA,
capacitive sensors will provide the readout of the relative position of
the satellites with respect to the freely 
flying test masses, which serve both
as interferometry end mirrors and drag-free orbit references (Bender,
2000). A drawback of electrostatic sensors is that the combined needs of
high precision ($\approx$ nm\rthz) and low electrostatic force
gradients require a small distance (or gap, $d$) between the test mass
and sensing electrodes. This need for proximity, introduces a number of
less easily characterized short range force effects, electrostatic and
otherwise, which grow with decreasing gap and can dominate the low
frequency test mass acceleration noise. An important example is DC
electric fields, which produce both force gradients (or ``stiffness'')
increasing as $d^{-3}$ and, coupled with charging and dielectric noise,
force noise proportional to, respectively, $d^{-1}$ and
$d^{-3/2}$. Though the current sensor design for the European LISA
demonstrator flight experiment LTP (Vitale, 2002), calls
for relatively large 4 mm gaps along the sensitive \x\ 
axis\footnote[1]{The test masses in both LISA and LTP have a single 
preferred measurement axis, referred to here as \x, 
in which it is essential to minimize the 
stray force disturbance.} in order to
limit these short range effects (Weber, 2002), DC fields
are still expected to be a significant noise source.

\begin{figure}
\centering
\includegraphics[width=110mm]{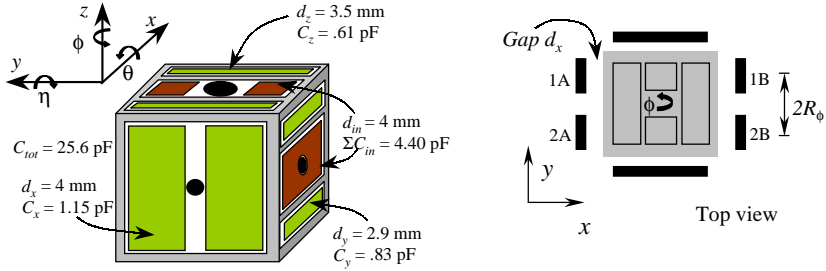}
\caption{Schematic of proposed capacitive sensor, with capacitance
and gap values, adapted from Weber (2002). The injection
electrodes used to provide the sensing bias are darkly shaded, the
sensing electrodes are medium gray, and grounded guard ring surfaces
are light gray. The planned test mass is a 46 mm, 2 kg Au / Pt cube.}
\label{fig_sensor}
\end{figure}

Stray DC fields, related to spatially varying DC surface potentials
known as patch fields, can arise from the different work function of
domains exposing different crystalline facets. These fields
statistically average over the small grains, which for gold surfaces 
can be micron size and produce RMS surface potential variations
of order 1 mV on mm length scales (Camp, 1992).
These are not likely to be a significant problem for the designed
sensor, shown schematically in \figr{fig_sensor}, where 4 mm \x\ axis
gaps provide distance for the fields and gradients to fall off, as well
as an effective dilution of the smaller length scale variations
(Speake, 1996). Potentially more dangerous are 
patch fields with relatively large coherence lengths, caused by
surface contamination from the assembly and from material
outgassing over a long mission. Noise models have assumed typical 
average whole electrode DC biases \dv{} of order 100 mV 
(Weber, 2002). 

Electrostatic actuation circuitry, to be integrated with the sensor for
application of forces using audio frequency modulated voltages, will
also allow direct application of DC voltages to the sensing electrodes.
We analyze here, as a way to reduce the total acceleration noise caused
by stray DC biases, the use of applied actuation voltages to first
measure the average biases and then compensate 
to make the average DC potential zero. The electrostatic model
considered here considers each electrode electrode at a single
potential, without spatial variation. This much simplified analysis
addresses what we can actually change with a single applied voltage per
electrode, the average DC potential. This approach is still relevant, as
the likely dominant effect, the interaction of DC fields with the noisy
test mass charge, is realistically parametrized, and thus curable, by
the average DC potential on the electrodes. We start with a description
of the noise sources related to stray DC biases, then turn to the
techniques for measuring and balancing them, both in flight and on the
ground. 

\section{NATURE OF THE PROBLEM: NOISY FORCES ARISING FROM DC FIELDS}
We analyze here the simplified model of the sensor in
\figr{fig_bias_meas}. Four \x\ sensing electrodes, opposing pairs 1A/1B
and 2A/2B, face the test mass along the \x\ axis. Functionally,
differential measurements of the two capacitor pairs are combined to
yield the test mass \x\ translational and $\phi$ rotational
displacements (Weber, 2002). Assigning electrode potentials $V_{j}$
and an accumulated test mass charge $q$, the instantaneous \x\ force
component $F_x$ and test mass potential $V_M$ can be expressed 
\begin{eqnarray}
F_x & = & \frac{1}{2}
\sum_{i}{\cder{i}{x} \left( V_i - V_M \right) ^2} 
\label{eqn_fx}
\\
V_M & = & \frac{q}{C_{T}} + \frac{1}{C_T}\sum_{j}{C_j V_j}
\label{eqn_vm}
\end{eqnarray}
$C_T$ is the total capacitance of the test mass to all
other surfaces. 
We sum here over all sensor conducting surfaces, $i$, but also 
note that \cder{i}{x}\ will have non-zero contributions only from the 4
\x\ sensing electrodes and, when the mass is translated off center in \x, 
from grounded guard ring surfaces (see \figr{fig_sensor}).

While the \x\ electrodes will nominally be held at DC ground by the
sensing / actuation circuitry (Weber, 2002), we consider here for each a
non-zero $V_i$ caused by the sum of a stray DC bias \dv{i}\ and a
thermal noise voltage $v_{ni}$ (later, we will consider the possibility
of compensating \dv{i}\ with an added applied DC bias, $V_{ai}$). Noisy
low frequency forces along \x\ arise in the interaction between \dv{i}\
and low frequency fluctuations in $q$ and $v_{ni}$. The charge $q$
accumulated from cosmic ray events evolves in a random walk
process, with an estimated effective step rate of $\lambda_{eff} \approx
260 $ $e$/s (Ara\'{u}jo, 2002), producing a ``red'' charge spectral
density $S_Q^{1/2} = \frac{e}{\omega} \sqrt{2 \lambda_{eff}}$. The
thermal noise $v_{ni}$, originating in lossy dielectric layers on the
electrode and test mass surfaces, is characterized by loss angle,
$\delta$, assumed here to be of order 10$^{-5}$ for Au coated surfaces
(Speake, 1999), with spectral density $\Sh{v_{n}} = \sqrt{4 k_B T
\frac{\delta}{\omega C_x}}$.

In evaluating Eqns. \ref{eqn_fx} and \ref{eqn_vm} for the force
disturbance, we define the nominal (test mass centered) capacitances and
first derivatives, $C_{x}$ and $\pm \frac{C_{x}}{d_x}$ (neglecting edge
effects, $d_x$ is the electrode - test mass gap). Substituting \dv{i},
$q$, and $v_{ni}$ into \eqr{eqn_fx}, we evaluate the leading order
force, and resulting acceleration noise spectral density \Sh{a}, caused
by the charge noise:
\begin{eqnarray}
 F_x & \approx &  
- \frac{q}{C_T} \frac{C_x}{d_x} \D{x}
\\ 
 \Sh{a_{charge}} & \approx &  
7 \times 10^{-15} \: \mathrm{m/s^2/\sqrt{Hz}} \times 
\left( \frac{\Delta_x }{\sqrt{4} \times 100 \: \mathrm{mV}} \right)
\left( \frac{\lambda_{eff}}{260 \: \mathrm{/s}}  \right)^{1/2}
\left( \frac{0.1 \: \mathrm{mHz}}{f} \right)
\: \: \: ,
\label{eqn_rand_charge}
\end{eqnarray}
where we define 
$\D{x} \equiv \left( \dv{1B} + \dv{2B} - \dv{1A} - \dv{2A} \right)$.
For the dielectric loss noise:
\begin{eqnarray}
 F_x & \approx &
%\frac{C_x}{d_x} 
%\left[ - v_{n1A} \dv{1A} +  v_{n1B} \dv{1B} 
%- v_{n2A} \dv{2A} + v_{n2B} \dv{2B} \right]
\sum_{i}{\cder{i}{x} v_{ni} 
\left( \dv{i} - \frac{q}{C_T} - 
\frac{1}{C_T} \sum_{j}{\dv{j} C_j}  \right)}
\\
 \Sh{a_{diel}} & \approx & 5 \times 10^{-16} \: \mathrm{m/s^2/\sqrt{Hz}} 
\times \left( \frac{\dv{}} {100 \: \mathrm{mV}} \right)
\left( \frac{\delta}{10^{-5}}  \right)^{1/2}
\left( \frac{0.1 \: \mathrm{mHz}}{f} \right) ^{1/2}
\label{eqn_diel}
\end{eqnarray}
To estimate a number for \Sh{a_{diel}}, 
we use $q = \sum_{j}{C_j \dv{j}} = 0$.  

The random charging produces a coherent disturbance
multiplied by the average difference
of DC potential seen on either side of the test mass, \D{x}. This is
true, neglecting edge effects, independently of the details of the
surface potential distribution. On the other hand, no such averaging
occurs in the ``beating'' of the DC voltages against the dielectric loss
thermal noise, which will be uncorrelated between electrodes. 

The acceleration noise target for LISA is $3\times 10^{-15}$
m/s$^2$\rthz, so the random charge noise in \eqr{eqn_rand_charge} is
potentially performance-spoiling at .1 mHz, the nominal low end of the
LISA measurement band and has an increasingly dominant effect with
decreasing frequency, where there is growing scientific interest to 
extend the gravitational wave measurement band (Bender, 2002). 
The dielectric loss
effect looks less threatening, but is difficult to estimate of the
contamination dependence of the loss angle $\delta$. With DC voltages
thus problematic at the 100 mV level, the effects described in Eqns.
\ref{eqn_rand_charge} and \ref{eqn_diel} illustrate the importance of AC
carrier voltages, rather than DC, in applying the necessary DC and low
frequency actuation forces, which demand RMS levels of several volts
(Weber, 2002).

\section{MEASUREMENT OF DC BIASES}
\label{sect_measure}
\begin{figure}
\centering
\includegraphics[width=105mm]{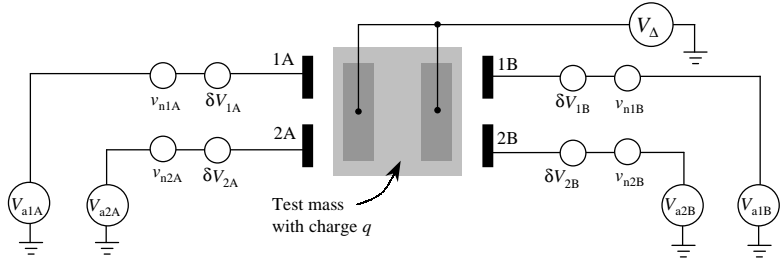}
\caption{Schematic of the simplified, four electrode electrostatic model
and measurement technique analyzed here. Stray DC biases, dielectric
noise, and test mass charge are denoted \dv{i}, $v_{ni}$, and $q$. In the
main measurement described in the text, a modulated
bias $V_{\Delta}$ is applied to the test mass via the \z\ electrodes.
Actuation voltages $V_{ai}$ can be applied to compensate the \dv{i} and can
also modulate the electrode voltages in other measurements.}
\label{fig_bias_meas}
\end{figure}

\begin{figure}
\centering
\includegraphics[width=105mm]{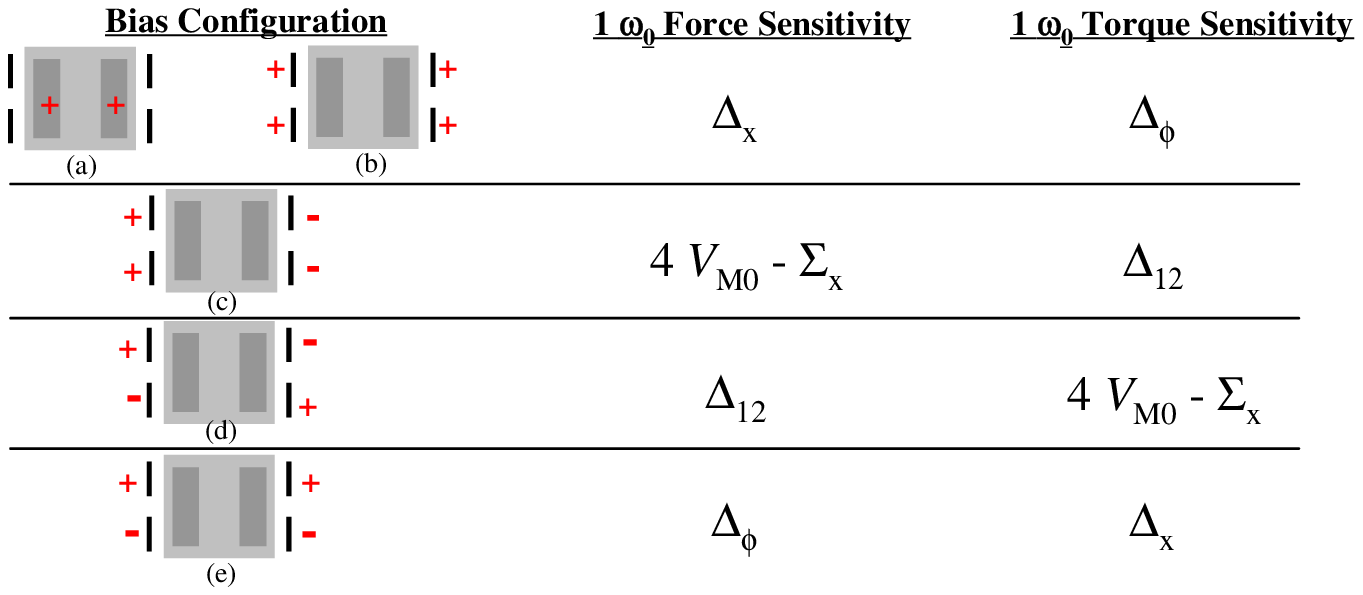}
\caption{Cartoon illustrating different electrode bias 
configurations useful in measurement of stray DC biases and charge, with
the specific DC bias 
combinations producing 1\wm\ forces and torques listed on the
right.  The + and - signs correspond to the relative phase of the modulated
voltage applied to a given electrode. }
\label{fig_meas_configs}
\end{figure}

There are a number of ways to measure the stray DC electrode biases.
Here we present one measurement in detail, illustrated in
\figr{fig_bias_meas} and (a) in \figr{fig_meas_configs}, and mention
several other, (b)-(e) in \figr{fig_meas_configs}. In configuration (a),
a sinusoidal signal $V_{\Delta} \sin{\omega_0 t}$ applied to the four
\z\ sensing electrodes makes the test mass potential oscillate with
amplitude $\frac{4 C_z}{C_T} V_{\Delta} \equiv \alpha V_{\Delta}$
($\alpha \approx .1$ for the sensor design in \figr{fig_sensor}). The
modulating test mass bias simulates, from the standpoint of forces in
the $xy$-plane, an oscillating test mass charge.

In expanding Eqns. \ref{eqn_fx} and \ref{eqn_vm} to first order in
displacement for the force $F_x$ (and an analogous equation for the
torque $N_{\phi}$), we make use of the second derivative $\cdder{x}{x}
\approx + \frac{2C_x}{d_x^2}$, as well as the derivatives involving the
mass rotation $\phi$, $\cder{x}{\phi} \equiv \pm
C_{x}\frac{R_{\phi}}{d_x}$, $\cdder{x}{\phi} \equiv + \frac{2C_x
R_{2\phi}^2}{d_x^2}$, and $\dcdxdphi\equiv \pm \frac{2C_x
R_{\phi}}{d_x^2}$ ($R_{\phi}$ is half the on-center electrode
separation, neglecting edge effects, and $R_{2\phi}$ has roughly the
same magnitude). The force and torque \footnote[2]{For simplicity, in
the torque formulae (Eqs. \ref{eqn_N_1f} and \ref{eqn_N_2f}), 
we omit the role of the \y\ electrodes,
which in this model would affect only the $\phi$ stiffness.} produced at
the first and second harmonic of the voltage modulation frequency
$\omega_0$ are given by
\begin{eqnarray}
\label{eqn_F_1f}
F_{1\omega_0} & = & \alpha V_{\Delta} \: \sin{\omega_0 t} \:  
\left\{ 
 - \D{x} \frac{C_x}{d_x} \: + \: 
\left[ \left( 4 V_{M0} - \Sigma_x \right) \frac{2 C_x}{d_x^2} 
+ V_{M0} \cdder{g}{x} \right] x \:
- \: \frac{2 C_x R_{\phi}}{d_x^2} \D{12}  \: \phi  \right\}
\\
 F_{2\omega_0} & = & 
- \alpha^2 V_{\Delta}^2 \cos{2 \omega_0 t}
\left( \frac{2 C_{x}}{d_x^2} + \frac{1}{4}  \cdder{g}{x} \right) \: x \: 
\label{eqn_F_2f}
\\
\label{eqn_N_1f}
 N_{1\omega_0} & = & \alpha V_{\Delta} \: \sin{\omega_0 t} \:
\left\{ - \D{\phi} \frac{C_{x} R_{\phi}}{d_x}  \: + \: 
\left[ \left( 4 V_{M0} - \Sigma_x \right) \frac{2 C_x R_{2\phi}^2}{d_x^2}
 + V_{M0} \cdder{g}{\phi} \right] \phi \: - \:
\D{12} \frac{2 C_x}{d_x} \: x
\right\}
\\
 N_{2\omega_0} & =  & - \alpha^2 V_{\Delta}^2 \cos{2 \omega_0 t} \: 
\left( \frac{2 C_{x} R_{2 \phi}^2}{d_x^2} + 
\frac{1}{4} \cdder{g}{\phi} \right) \: \phi \: 
\label{eqn_N_2f}
\end{eqnarray}
For convenience we have defined the combined stray DC biases $\D{\phi}
\equiv \left( \dv{1A} + \dv{2B} - \dv{2A} - \dv{1B} \right)$,$\D{12}
\equiv \left( \dv{2A} + \dv{2B} - \dv{1A} - \dv{1B} \right)$, and
$\Sigma_x \equiv \left( \dv{1A} + \dv{1B} + \dv{2A} + \dv{2B} \right)$.
We also make use of the test mass potential when centered in the absence
of any applied fields, $V_{M0} \equiv \frac{q}{C_T} +
\frac{\sum_{j}{\delta V_j C_{j0}}}{C_T}$.

The 2\wm\ terms, in Eqns. \ref{eqn_F_2f} and \ref{eqn_N_2f} are pure
``stiffness,'' or position dependent, terms related only to the applied
bias. The test mass can be positioned to make the 2\wm\ force 
signal disappear, which centers the mass to
then eliminate the stiffness terms in the 1\wm\ force and torque
terms, making the measurement independent of $V_{M0}$ (and charge). 
The self-calibrating location of the ``force zero'' with
the 2\wm\ force signals is useful because the force zero
(roughly speaking, where the \emph{capacitance derivatives} are equal)
will not, due to machining imperfections and asymmetries, coincide
exactly with the sensor readout zero, where the \emph{capacitances}
themselves are equal. With the test mass centered, the remaining \x\
force is thus a 1\wm\ signal cleanly proportional to the differential
bias \D{x}\ of relevance to the random charging force noise. Likewise, a
null measurement test is possible after compensation voltages are
applied. 

In principle, the forces and torques detectable in Eqns. \ref{eqn_F_1f}
- \ref{eqn_N_2f} offer all the information measurable for the average
electrode DC biases and test mass charge. \D{x}\ and \D{\phi}\ are the
main 1\wm\ signals in force and torque, with \D{12}\ and $\left(
\Sigma_x - 4 V_{M0} \right)$ entering in the stiffness and thus
measurable in the position dependence of the signals. These quantities
can, however, be measured better individually, with the test mass
centered, using opportune modulation of the channel 1 and 2 actuation
voltages. Four symmetric schemes, which permit the 2\wm\ centering
procedure and together give a complete average DC bias characterization
by measurement of the 1\wm\ force terms, are illustrated schematically
in \figr{fig_meas_configs} (b) - (e).

It is worthwhile commenting here on the applicability of this simplified
model, given the realistic spatial fluctuations present in electrode
surface potentials. If we considered \x\ electrodes divided into many
surface domains, each with a different bias \dv{i} but all still a
distance $d_x$ from the test mass, and apply the configuration (a)
$V_{\Delta}$ bias, the centering process with the 2\wm\ signal would be
unchanged and, neglecting edge effects, the resulting 1\wm\ force would
still give the average potential difference \D{x}. If we also included
variations in the ``grounded'' guard ring potential, the effective
average potential \D{x}\ relevant to the detected force would then also
average over the guard rings stray bias distribution. Importantly,
however, the applied biases $V_{ai}$ that null the force signal in this
measurement will also null the effect of random charge on the test mass,
including all stray DC fields biases generated inside (and even outside)
the sensor. 

The torques do not generalize as easily from the simple ``one voltage per
electrode'' picture to spatially variable potentials, because the
average potential of relevance to the torques is weighted by an
effective arm length, or distance from the rotation axis. 
Thus, to balance the average electrode potentials relevant
to the \x\ force noise, the additional differential bias 
potentials are best measured by the \x\ forces excited by 
the bias configurations in \figr{fig_meas_configs}. The configurations
using the \x\ electrodes themselves for biasing are sensitive to the
biases on those electrodes, much less so to any guard ring patch
effects. Combining the measured \D{x}, \D{\phi}, \D{12}, and $\left( 4
V_{M0} - \Sigma_x \right)$ allows calculation of the potentials $V_{ai}$
needed to null the detected forces and thus to bring all four sensing
electrodes near to the test mass potential.

This procedure can be repeated for electrodes on the \y\ and \z\
faces. We finally note that the charge measurement configurations, such as
(c) in \figr{fig_meas_configs}, measure not just the charge but $\left(
4 V_{M0} - \Sigma_x \right)$ and thus will depend quantitatively 
on how well the surrounding DC biases are balanced.

\section{IN-FLIGHT DETECTION AND BALANCING}
\label{sect_flight}
In flight, the forces and torques described above can be detected by the
position sensor signal itself. Both on LISA and LTP, a low frequency
electrostatic suspension mode will be used to control the test mass
non-drag-free degrees of freedom, and can also be applied on the
sensitive \x\ axis to make this measurement (Bortoluzzi, 2002). The
suspension will have high gain at low frequencies to counteract DC
forces and very low gain in the measurement band, such that, for forces
applied at mHz frequencies, the test mass dynamics is roughly that of a
free particle, with displacement $x \approx \frac{-f}{m\wsq{}}$. For a 1
mHz modulation of $V_{\Delta} = 1 V$ (and thus roughly 100 mV test mass
bias amplitude), the displacement would be of order 40 nm if \D{x}
$\approx$ 100 mV.

The measurement resolution will likely be dominated by sensor noise and
residual spacecraft motion, rather than the force noise acting on the
test mass. For the LTP experiment, the goal for this total
position noise is 5 nm\rthz (Bortoluzzi, 2002). The resulting resolution
of \D{x}\ in integration time $T$ is then
\begin{equation}
\Delta \left( \D{x} \right)
\approx \frac{1}{ \left| \frac{\partial F}{\partial \D{x}} \right|}
\frac{\sqrt{2} m \wsq{0} \Sh{\x_n}}{\sqrt{T}}
\approx 350 \, \mathrm{\mu V} \: \times 
\left( \frac{1 \, \mathrm{hr}}{T} \right)^{1/2} 
\left( \frac{1 \, \mathrm{V}}{V_{\Delta}} \right) 
\left( \frac{\Sh{\x_n}}{5 \, \mathrm{nm/Hz^{1/2}}} \right)
\left( \frac{\omega_0}{2\pi \times 1 \, \mathrm{mHz}} \right)^2 
\label{eqn_res_dvphi}
\end{equation}

A one hour experiment could thus allow sub-mV resolution of the stray DC
bias difference \D{x}. The resolution limit in the balancing of the
applied biases will be limited, for LTP, to the 1 mV level by the
envisioned 16 bit, $\pm$ 40 V full scale actuation DA converter and
amplifier (Vitale, 2002). Any instability in the balancing voltage will
multiply this residual mV level voltage to produce a noisy force,
but the projected actuation voltage stability, of order $10^{-6}$\rthz,
makes this a neglible source of force noise (Vitale, 2002). Note that
balancing of \D{x} to the mV level would reduce the acceleration noise
caused by random charging below the .1 fm/s$^2$\rthz\ level at .1 mHz,
and thus well below the LISA target acceleration noise. 

\section{TORSION PENDULUM GROUND TESTING}
Current ground based measurement programs will use torsion pendulums to
characterize weak forces and force gradients of relevance to drag-free
control (Hueller, 2002). Suspending a test mass as the pendulum inertial
member, surrounded by a LISA or LTP prototype sensor, permits sensitive,
high isolation, low restoring force measurements for a single torsional
degree of freedom. A pendulum comprised of a single test mass suspended
on axis with the torsion fiber is sensitive to torques, while a pendulum
with a test mass held off the rotation axis will convert translational
forces into measurable pendulum angular deflections.

With the single mass configuration, the rotational DC bias imbalance
\D{\phi} can be measured using a \z\ electrode bias $V_{\Delta}$, with
the torque (see \eqr{eqn_N_1f}) detected in the pendulum deflection. The
measurement resolution will likely be limited by the mechanical torque
noise acting on the pendulum, \Sh{N}, 
\begin{equation}
\Delta \left( \D{\phi} \right)
\approx \frac{1}{ \left| \frac {\partial N}{\partial \D{\phi}} \right |}
\frac{\sqrt{2} \Sh{N}}{\sqrt{T}}
\approx 400 \, \mathrm{\mu V} \: \times 
\left( \frac{1 \, \mathrm{hr}}{T} \right)^{1/2} 
\left( \frac{1 \, \mathrm{V}}{V_{\Delta}} \right) 
\left( \frac{\Sh{N_n}}{5 \, \mathrm{fN\,m/Hz^{1/2}}} \right)
\label{eqn_res_tor}
\end{equation}
Here, 5 fN\,m\rthz\ refers to the expected 1 mHz
thermal torque noise limit for the designed one mass pendulum
(Hueller, 2002).  Higher torque noise would necessitate an increased 
$V_{\Delta}$ to maintain this resolution.
  
The torsion pendulum measurement has several possible systematic errors.
Because the pendulum is relatively insensitive to purely translational
forces, there will not be a 2\wm\ force signal to center in \x.
Centering in \x\ will thus depend on the sensor translational readout
zero, which could differ from the force zero by order 10 $\mu$m, set by
the sensor machining tolerances (Vitale, 2002). The offset in \x\ means
the \x\ stiffness term in \eqr{eqn_N_1f} will not vanish, producing, for
$\left| \D{12} \right| \approx $ 200 mV, a systematic error of roughly 1
mV in the measurement of \D{\phi}. Another error enters in the finite
``twist-tilt'' feedthrough (Smith, 1999) that can couple the
electrostatic force, proportional to \D{x}, into the measured torsional
rotation. However, assuming that \D{\phi}\ and \D{x}\ are of the same
order of magnitude, the ratio between the excited pendulum swing angle
and the torsional twist angle will be of order 10$^{-6}$, giving an
immeasurably small effect for typical values of twist-tilt coupling.
Thus, measurement and balancing of $\D{\phi}$ looks feasible at the mV
level with the torsion pendulum and could provide a valuable ground test
of the compensation technique suggested here. 

\section{CONCLUSION}
The true magnitude and long term stability 
of the stray DC potentials for LISA will not be known until the actual
flight experiment begins. However, the average fields are
measurable and can be balanced. Compensating the average stray electrode
biases should be quite effective in cancelling a potentially large
random charging noise source, where balancing the average potential seen
on either side of the test mass should null the force from a deposited
charge. For the lossy dielectric noise, however, which itself could be
spatially dependent, the average potential alone may not adequately
characterize the phenomena, and the success of the compensation in
reducing this source of acceleration noise will depend on the extent to
which the thermal noise causes the ``patches'' across an entire
electrode to fluctuate in unison. 

\section{ACKNOWLEDGEMENTS}
The authors would like to acknowledge Peter Bender for bringing the 
the random charging problem, and its 
possible compensation, to our attention.

\section{}
Corresponding author: W. J. Weber, \emph{weber@science.unitn.it}
\end{document}